\documentclass[twocolumn,secnumarabic,amssymb, nobibnotes, aps,prl,reprint,letterpaper,superscriptaddress]{revtex4-2}
\usepackage{float}

\usepackage{amsmath}
\usepackage{amssymb}
\usepackage{physics}
\usepackage{graphicx}
\usepackage{siunitx}

\newcommand{\figref}[1]{Fig.~\ref{#1}}

\begin{document}
\title{Quantum interference between spectral bandwidth mismatched photons}
\author{Jan Krzy\.zanowski}
\email{jan.krzyzanowski@fuw.edu.pl}
\affiliation{Faculty of Physics, University of Warsaw, Pasteura 5, 02-093 Warszawa, Poland}
\author{Jerzy Szuniewicz}
\affiliation{Faculty of Physics, University of Warsaw, Pasteura 5, 02-093 Warszawa, Poland}
\author{Sanjay Kapoor}
\affiliation{Faculty of Physics, University of Warsaw, Pasteura 5, 02-093 Warszawa, Poland}
\author{Filip So\'snicki}
\affiliation{Faculty of Physics, University of Warsaw, Pasteura 5, 02-093 Warszawa, Poland}
\affiliation{Integrated Quantum Optics, Institute for Photonic Quantum Systems (PhoQS), Paderborn University, Warburger Str.~100, D-33098, Paderborn, Germany}
\author{Micha\l{} Karpi\'nski}
\email{mkarp@fuw.edu.pl}
\affiliation{Faculty of Physics, University of Warsaw, Pasteura 5, 02-093 Warszawa, Poland}

\begin{abstract}

Two-photon interference is a cornerstone of photonic quantum technologies. However, its practical use in hybrid architectures is constrained by the need for photon indistinguishability, particularly in linewidth and the associated time scale. While narrowband filtering can improve interference visibility, it introduces significant photon loss~--~a critical limitation for applications. Here, we experimentally demonstrate an efficient approach to enable non-classical two-photon interference between spectral-bandwidth mismatched photons using an electro-optic time lens. We increase the visibility of Hong-Ou-Mandel (HOM) interference between spectral bandwidth mismatched photons from $(31.0\pm1.7)$\%, to $(91.7\pm2.6)$\%, with correction for accidental coincidences, achieving highly non-classical two-photon interference visibility without spectral filtering. Moreover, we demonstrate a simultaneous correction of the central wavelength mismatch of $\pm1.1$~$\si{nm}$, while maintaining the non-classical interference visibility. This result opens the possibility to efficiently integrate quantum systems operating at different time scales for hybrid quantum communication, teleportation, entanglement swapping, distributed sensing, and hybrid quantum computing.
\end{abstract}

\maketitle

Indistinguishable photons~--~photons identical in all degrees of freedom~--~are a fundamental resource for photonic quantum computing \cite{Knill2001scheme,Lu2007demonstration,Lanyon2007experimental,Politi2009shor,Lu2018quantum,Wang2017high}, quantum metrology \cite{Walther2024}, quantum communication \cite{Lo2012measurement}, and in hybrid quantum networks \cite{Main2015distributed}. Applications, such as long-distance entanglement distribution based on quantum repeaters \cite{Duan2001long,Sangouard2011quantum} rely on two-photon interference of photons originating in two separate photon sources. Hybrid two-photon interference is an inherently challenging task, as various sources emit photons with a wide variety of central wavelengths and spectral-temporal shapes, limiting the interference visibility  \cite{polyakov2011,huber2017interfacing,kim2026two}. While in recent years tremendous progress has been made in sources of highly indistinguishable single photons realized via, e.g., nonlinear optical crystals \cite{graffitti2018independent,pickston2021optimised}, or single quantum emitters \cite{gazzano2013bright,somaschi2016near}, the challenge remains as different applications require identical spectral-temporal properties of single photons. Therefore, one has to ensure that photons from different sources are spectrally and temporally matched. The central wavelength matching can be solved via quantum frequency conversion \cite{mann2025hong,PhysRevA.88.042317,ates2012two,weber2019two,wright2017spectral,kapoor2025electro} for significant mismatches, while fine tuning and spectral bandwidth matching requires a different approach.

The obvious way to improve indistinguishability of spectrally overlapping photons is the application of a spectral filter to the spectrally broader photon. This technique has been shown to improve the visibility of two-photon interference between mismatched sources, such as a heralded spontaneous parametric down conversion (SPDC) photon and a quantum dot photon \cite{huber2017interfacing,polyakov2011}. Its major drawback is the inherent high loss introduced by rejecting the photons that do not match the narrow filter linewidth. To address this issue, spectral shaping techniques have been developed that do not involve filtering. They utilize nonlinear \cite{lavoie2013spectral,allgaier2017highly,joshi2022picosecond} or linear (electro-optic) \cite{karpinski2017bandwidth,sosnicki2023interface,mittal2017temporal,zhu2022spectral,seidler2020spectral} optical methods, applying spectral and temporal phases to modify the spectral-temporal profile of single photons. However, the indistinguishability of photons after such modifications was verified experimentally for spectral shifts only \cite{ates2012two,weber2019two,wright2017spectral,kapoor2025electro,fan2016integrated}. The phase-only spectral bandwidth  modification techniques, which improve spectral bandwidth indistinguishability \cite{Horoshko2023PhysRevA,Horoshko2023Optica} have been applied to single photons and photon pairs \cite{lavoie2013spectral,allgaier2017highly,karpinski2017bandwidth,sosnicki2023interface,mittal2017temporal,zhu2022spectral,mazelanik2022optical,joshi2022picosecond,fan2019spectrotemporal,myilswamy2020spectral,seidler2020spectral,couture2025terahertz}, but their two-photon interference has not yet been demonstrated.

\begin{figure}[!b]
\centering
\includegraphics[width=\columnwidth]{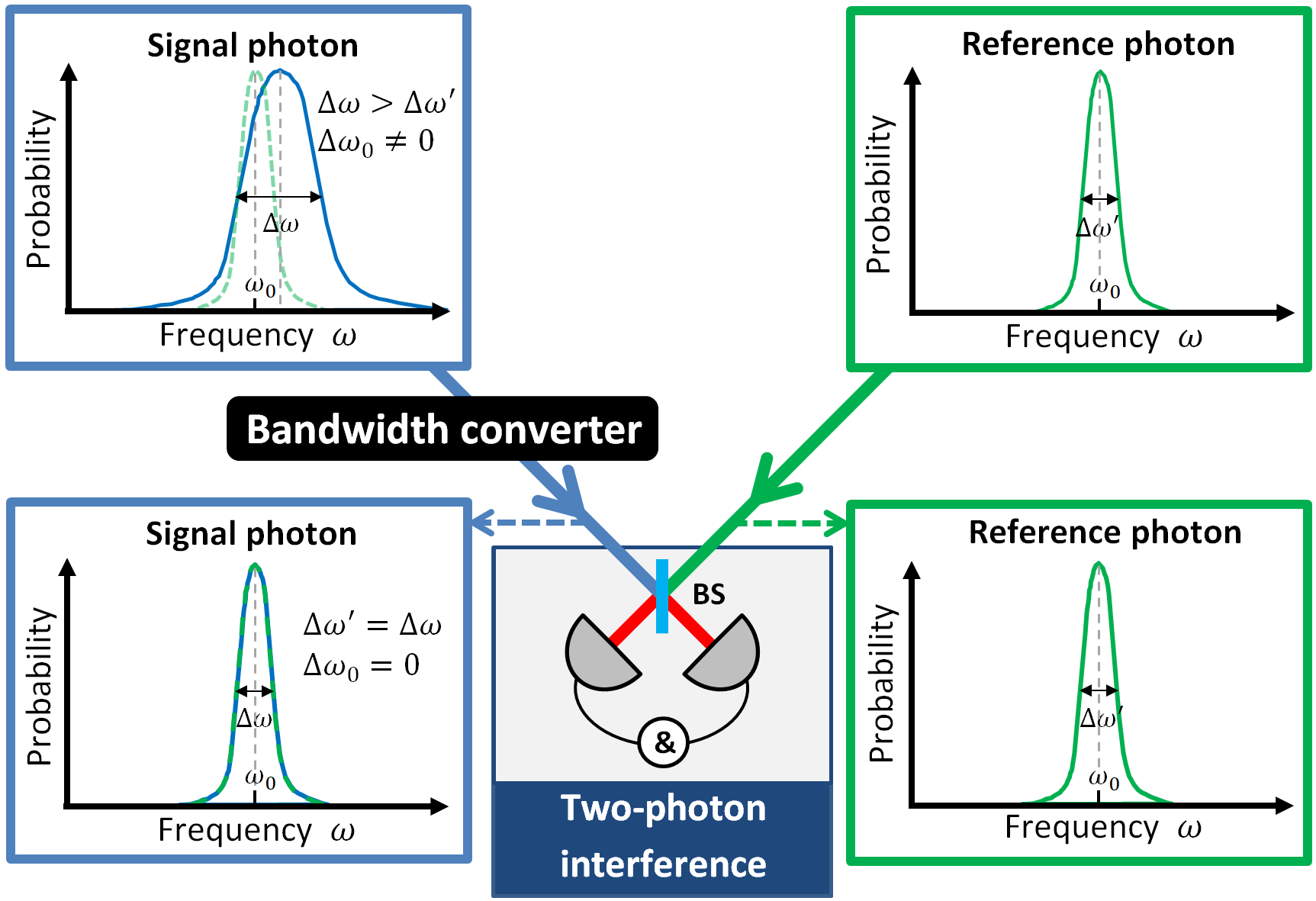}
\caption{Overview of the experiment. We begin with two photons with mismatched spectral bandwidths. We apply a time-lens-based spectral bandwidth converter to compress and shift the spectrum of the signal photon to match the spectral width and central wavelength of the reference photon. Such modification improves the overlap of the photons' spectral-temporal modes and thereby improves the two-photon interference visibility. \label{fig:Overviev}}
\end{figure}

In this work, we demonstrate non-classical two-photon interference between photons with initially mismatched bandwidths and central wavelengths. By using a spectral bandwidth converter based on an electro-optic time lens, we enhance their spectral indistinguishability. This converter rearranges the spectral components, producing a narrower spectrum without spectral filtering.

We present a conceptual scheme of our experiment in \figref{fig:Overviev}. We start with two photons with mismatched spectral bandwidths. One photon has a much broader spectrum than the other, significantly limiting their interference visibility. We apply an electro-optic time-lens-based spectral bandwidth converter \cite{karpinski2017bandwidth,torres2011space} (see Supplementary Sec. S1 for details) to the photon with a broader spectrum (the signal photon),  compressing its spectrum to the approximate spectral width of the reference photon. We perform Hong-Ou-Mandel interference (HOMI) \cite{hong1987measurement} between the spectrally-compressed and reference photon to observe the visibility improvement.

\begin{figure*}[!t] 
\centering
\includegraphics[width=1.75\columnwidth]{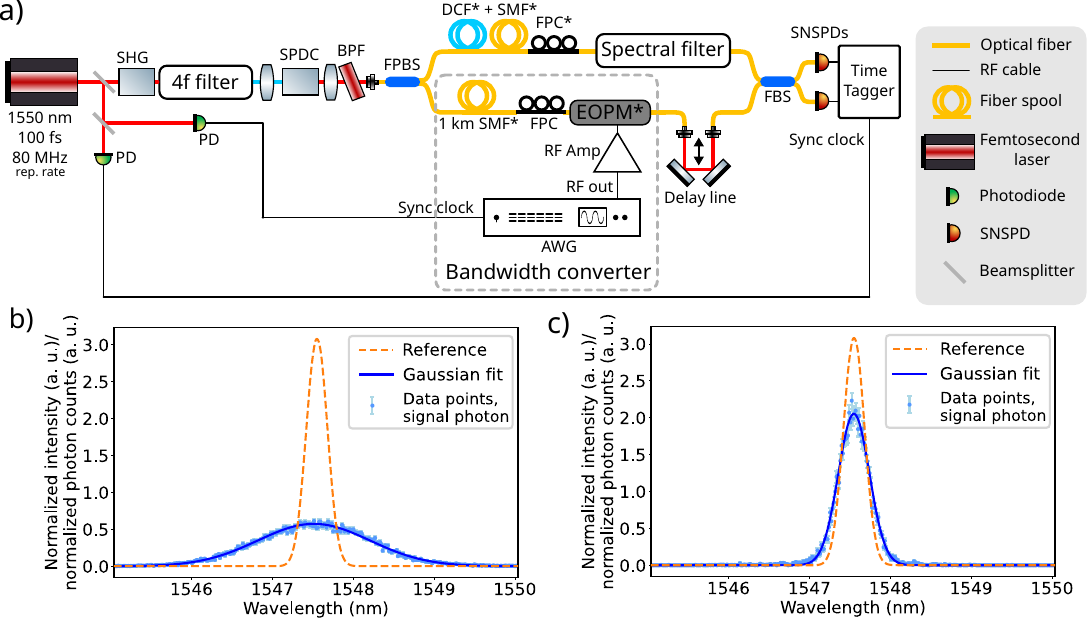}
\caption{Schematic of the experimental setup~(a). A pair of photons is generated by the SPDC source, filtered with a broadband bandpass filter (BPF). The reference photon is spectrally filtered using a $4f$ filter. The second (unfiltered) photon is spectrally compressed using the time-lens-based bandwidth converter. We drive the electro-optic phase modulator (EOPM) with an RF signal from an arbitrary waveform generator (AWG) amplified by an RF amplifier. We perform two-photon interference at a $50:50$ fiber beam splitter (FBS). Elements marked with an asterisk were removed from the experimental setup to measure the reference HOM dip without the bandwidth compression. (b,~c)~--~the action of the time lens on the wideband photon's spectrum. The single-photon spectra (data points) were acquired as  coincidence measurements with the filtered reference photon, solid lines: Gaussian fits to the data. The spectrum in~(b) is compressed by the converter resulting in the spectrum presented in~(c). The reference spectra (dashed line) were measured with classical bright light using the OSA. The FBS was not used for measurements in~(b),~(c). \label{fig:Spectra}}
\end{figure*}

We generated photon pairs using type-II SPDC in a periodically-poled potassium titanyl phosphate (PPKTP) waveguide, as shown in \figref{fig:Spectra}(a). The waveguide was pumped with second harmonic ($773.5$~$\si{nm}$ wavelength) of the output of the C-band erbium femtosecond oscillator and amplifier (Menlo Systems C-Fiber HP, $80$~$\si{MHz}$ repetition rate). The pump was spectrally filtered to $1.3$~$\si{nm}$ FWHM (full width at half maximum) to produce spectrally degenerate photon pairs centered around $1547.5$~$\si{nm}$. Both photons then pass through a 1.8 nm-FWHM bandpass filter that removes the sidelobes of the periodically poled phase-matching, leaving a spectrally separable biphoton state \cite{grice1998spectral}. The corresponding joint spectral intensity (JSI) is shown in Supplementary Sec. S5A.

The photon pairs are separated using a fiber polarization splitter, the signal photon is sent to the bandwidth converter, and the idler photon through a home-built $4f$ spectral filter. The reference photon is prepared by filtering the idler photon's spectrum from $1.8$~nm to $0.3$~nm FWHM (calibrated with classical light). Detuning the filter also shifts the reference photon's central wavelength across its $1.8$-nm-wide spectrum, allowing us to introduce an additional central-wavelength mismatch.

At this stage of the experiment the signal and reference photons exhibit different spectral bandwidths (see \figref{fig:Spectra}(b)). We recover their spectral compatibility by introducing the spectral bandwidth converter (see Supplementary Sec. S1 for more details) in the signal photon path. The bandwidth converter consists of: $0.5$~$\si{km}$ of passively thermally stabilized SMF-28 fiber to introduce $(10.7\pm0.1)$~$\si{ps^2}$ of group delay dispersion (GDD) followed by a fiber-coupled lithium-niobate-waveguide electro-optic phase modulator (EOPM, EO-Space, $16$~$\si{GHz}$ bandwidth, $3$~$\si{V}$ half-wave voltage @ $1$~$\si{GHz}$). We drive the EOPM using a $16$~$\si{GHz}$ sine waveform generated by an arbitrary waveform generator (AWG, Keysight M8196A). Around its extremum, the sinusoidal waveform approximates the quadratic phase required for the spectral compression \cite{torres2011space}. We synchronized it with the laser pumping the SPDC source, which resulted in a stable delay between the radio-frequency (RF) signal and the photon arrival (up to residual thermal drift of the fiber length). The waveform generated by the AWG was amplified using a $3$~$\si{W}$ RF amplifier (Mini Circuits ZVE-3W-183+). We adjusted the signal amplitude at the AWG to achieve the best visibility of HOM interference. The power transmission of the bandwidth converter is $28.1$\%, while the measured power transmission of the filter applied to the reference photon is only $7.6$\%. This value reflects only the spectral mismatch; insertion losses would further reduce the transmission of a real filter.

We show the spectral modifications introduced by the  converter in \figref{fig:Spectra}(b,c) and \figref{fig:ResultsPanel2}(b,d,f). The single-photon spectra are measured with a time-of-flight spectrometer utilizing the dispersive Fourier transform (DFT) technique \cite{avenhaus2009fiber}. As the DFT dispersive medium, we used $40.9$~$\si{km}$ (and $41.7$~$\si{km}$ for JSI) of SMF-28 fiber, which gives the spectral resolution of $0.15$~$\si{nm}$ (GDD-limited) \cite{goda2013dispersive}. The photons were detected by superconducting nanowire single photon detectors (SNSPD, Single Quantum EOS, 11 ps RMS timing jitter) and time tagged with a time-to-digital converter (Swabian Instruments Time Tagger Ultra). We measured the single-photon spectra shown in \figref{fig:Spectra}(b,c) and \figref{fig:ResultsPanel2}(b,d,f), while heralding the signal photon with the filtered reference photon. The reference spectrum in both figures is the transmission of the spectral filter ($0.3$~$\si{nm}$ FWHM) measured with classical light, using an optical spectrum analyzer (OSA, Yokogawa AQ6370D).

By applying the bandwidth converter we regain matching of the signal and reference photons' spectral FWHMs. Before the bandwidth converter (\figref{fig:Spectra}(b)) the spectral width (FWHM) of the signal photon was $(1.6\pm0.1)$~$\si{nm}$. After compression the measured FWHM decreased to $(0.4\pm0.1)$~$\si{nm}$ (\figref{fig:Spectra}(c) and \figref{fig:ResultsPanel2}(b)). The difference between the spectral widths measured with single-photons and classical light arises mainly from the drift of the delay between the single photon arrival and the RF signal as well as the uncertainty of the exact GDD value.

\begin{figure*}[!t]
\centering
\includegraphics[width=1.9\columnwidth]{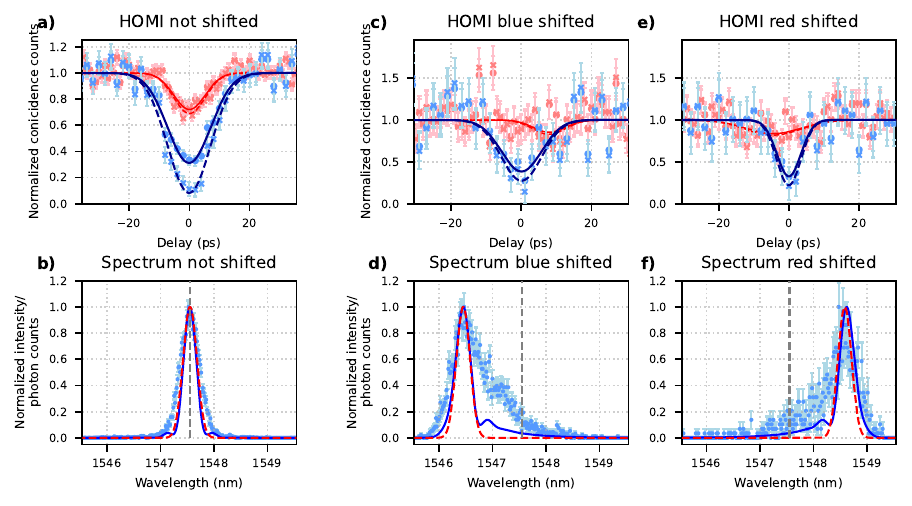}
\caption{HOM interference of spectrally mismatched photons: (a) normalized coincidence rates between beam splitter outputs as a function of delay between the interfering photons, set at the delay line. HOM dip after compressing the signal photon (blue), vs the dip without the bandwidth converter (red). We calculate the two-photon interference visibility from fitted Gaussian functions (solid lines). Error bars assume Poisson photon count statistics. With crosses and dashed lines we plot respectively data points and fits with correction for accidental coincidences (see sec. S5B). The zero point of the delay axis for the red and blue data was determined independently (see sec. S5C and Fig.~S8) due to the need to remove the components marked with asterisk in \figref{fig:Spectra}(a) to acquire the red data. (b) spectra of the compressed signal photon (experimental data points) measured in coincidence with the reference photon, solid lines: the spectral shapes predicted by the numerical simulation. We assumed central wavelength matching for the wavelength axis calibration. The reference spectrum (dashed line) represents the transmission window of the filter applied to the reference photon and was measured with classical bright light using the OSA. For (a) and (b) the filter is centered at the reference photon's central wavelength (grey dashed line). In (c,d) and (e,f) we show the data measured with the filter respectively blue and red shifted from the unmodified signal photon's central wavelength.
\label{fig:ResultsPanel2}}

\end{figure*}

After matching the bandwidths of the photons, we performed the HOMI measurement. We added $0.5$~$\si{km}$ of fiber to the reference photon's path ($58$~$\si{m}$ of dispersion-compensating fiber and $454$~$\si{m}$ of SMF-28, net GDD below $0.5$~$\si{ps^2}$) to match propagation delays without introducing additional distinguishability. We interfered the compressed signal photons and the reference photons in a fiber-based 50:50 beam splitter (FBS) and detected the coincidences between the photons at the FBS outputs via SNSPDs followed by a time-tagger. We applied a coincidence window of $2$~$\si{ns}$, which is shorter than the inverse of the repetition rate of the source ($12.5$~$\si{ns}$) but much longer than the duration of the photons (about $18$~$\si{ps}$).Since the HOMI dip is observed in the photon coincidence rate, the unequal average number of photons at each beam splitter input in principle does not affect the measured visibility \cite{hong1987measurement,prabhakar2020two,drago2024hong}.We used a fiber-coupled free-space delay line to control the relative delay between the photons. We measured the coincidence counts at each delay for $30$~$\si{s}$ with step of $2$~$\si{ps}$. 

The resulting HOM dip is shown in \figref{fig:ResultsPanel2}(a) in blue. We fitted a Gaussian function (with the width, position on delay axis, multiplicative and additive constants as fitting parameters with no constraints) to the coincidence counts, which resulted in a visibility of $(68.7\pm2.0)$\%, exceeding the non-classical threshold of $50$\% \cite{kim2013conditions}. We used the bootstrap method to estimate the uncertainty assuming Poissonian photon count statistics \cite{PhysRevX.12.031033,karpinski2017bandwidth}.

The large transmission difference between the converter and filter arms makes accidental coincidences a notable contribution to the measured visibilities. They originate from the finite extinction ratio of the polarizing beam splitter that separates the signal and idler: unsplit pairs traverse the single path (converter or filter) and produce accidental coincidences that are comparable in number to the true events. To reduce that problem, we intentionally decreased the converter's transmission to reduce the number of accidental events. We independently measured the accidental-coincidence rate and subtracted it in the HOMI analysis (Supplementary Sec. S5B).

We present the data with correction for accidental coincidences in \figref{fig:ResultsPanel2}(a) with a dashed line. The correction further increased the interference visibility to $(91.7\pm2.6)$\%, which is close to the numerically simulated value (see Supplementary Material for details).

We attribute the non-perfect interference visibility primarily to the low coincidence-to-accidental ratio resulting from high losses in the preparation of the narrowband reference photon by spectral filtering.  Additionally, non-optimal dispersion in our experimental setup reduces the maximum achievable visibility. We utilized a GDD of $\Phi = 10.7\pm0.1$~ps$^2$ instead of the optimal $\Phi \approx 8.3$~ps$^2$. Numerical simulations of this specific configuration, where the modulation amplitude was optimized to $A = (3.6\pm 0.1)\pi$ rad, confirm that the visibility is theoretically limited to $92.5\%$, and including $\pm1$~ps phase offset between the time lens and the photons, which translates to a residual central wavelength mismatch \cite{kauffman1994time}, results in an average visibility of $(88.4 \pm 5.6)\%$. Operating away from the optimal dispersion amplifies the higher-order aberrations of the sinusoidal phase, producing pronounced spectral side-lobes \cite{kapoor2025aberration} that introduce distinguishability. With the optimal dispersion and order of magnitude higher modulation amplitude ($\approx57 \pi$) a visibility exceeding $99$\% would be achievable. Details of these simulations along with analytical calculations are presented in the Supplementary Material.

To show the advantage of our bandwidth conversion scheme we removed the bandwidth converter and the reference photon's fiber delay from the setup and measured the HOM dip shown in \figref{fig:ResultsPanel2}(a) in red. For such highly mismatched photons with the spectra presented in \figref{fig:Spectra}(b), the dip has a much lower visibility of $(31.0\pm1.7)$\% (Gaussian fit) with correction for accidental counts ($(28.0\pm1.6)$\% without accidentals correction).

We further analytically estimated the advantage of spectral compression in a practical scenario of the realization of quantum teleportation \cite{bouwmeester1997experimental} using HOM interference with a bandwidth converter as a Bell-state measurement (BSM). The success rate of HOM-based quantum teleportation is proportional to $T_{\mathrm{conv}} \cdot V_{\mathrm{HOM}}$ \cite{bouwmeester1997experimental,wang2015quantum}, where $T_{\mathrm{conv}}$ is the transmission of the bandwidth converter arm. For the time-lens converter, $T_{\mathrm{lens}} V_{\mathrm{lens}} = 0.281 \times 0.917 = 0.258$, where $V_{\mathrm{lens}}$ is the HOM visibility from our experiment. The reference filtering scheme, using the same spectral filter, reaches at best $T_{\mathrm{filter}} V_{\mathrm{filter}} = 0.076 \times 1 = 0.076$, assuming perfect visibility of HOM interference ($V_{\mathrm{filter}} = 1$). The bandwidth converter therefore yields approximately a 3.4× improvement in the rate of correctly teleported states, including its insertion loss.

To demonstrate simultaneous matching of the central wavelength and the spectral bandwidth, we repeated the HOM measurement with the reference-photon filter detuned by $\pm1.1$~nm from the central wavelength of the signal photon. The converter must then compress the signal spectrum and shift its central wavelength at the same time, combining the bandwidth compression with the central wavelength shifting demonstrated by Wright et al. \cite{wright2017spectral} in a single time-lens setup. We introduce a temporal detuning between the sinusoidal phase and the photon's arrival time so that the photon also samples the linear part of the sinusoid, which imparts the central-wavelength shift \cite{wright2017spectral}; enough quadratic curvature remains to compress the spectrum while simultaneously performing blue- or red-shift of the spectrum, as shown in \figref{fig:ResultsPanel2}(d,f).

The resulting HOM dips are shown in \figref{fig:ResultsPanel2}(c,e). To counteract the increased accidental coincidence rate, we added fiber attenuator to the converter setup to reduce the discrepancy between the rates. From the raw data, we obtain visibilities of $(62\pm10)\%$ and $(67\pm12)\%$ for the blue- and red-shifted reference filter, respectively, increasing to $(73\pm13)\%$ and $(78\pm16)\%$ after correction for accidental coincidences~--~both well above the $50\%$ non-classical threshold. The corrected values agree well with simulations (see Supplementary Material). The visibility is lower than for the unshifted case because of temporal aberrations \cite{kapoor2025aberration}. The same detuning that produces the central-wavelength shift moves part of the temporally dispersed wavepacket beyond the region where the sinusoid approximates a quadratic phase~--~that is, outside the effective temporal aperture of the time lens~--~such that these components are imperfectly compressed and distort the converted spectrum, reducing its overlap with the reference photon.

The reference HOM scans, measured without the bandwidth converter and plotted in red in \figref{fig:ResultsPanel2}(c,e) do not show clearly visible dips. The Gaussian fits yield the visibilities of $(17\pm14)$\% and $(18.2\pm8.4)$\% (with correction for accidental coincidences) for the blue- and red-shifted filter respectively ($(15\pm12)$\% and $(16.4\pm7.6)$\% without correction). However, the position of the fitted dips relative to the zero delay (see Sec. S5C) indicates that an accurate fit may be precluded by the expected low visibility combined with measured low coincidence rates.

In summary, we observed non-classical two-photon interference of spectrally mismatched photons. Our experimental results show that introducing a time-lens-based spectral bandwidth converter improves HOM visibility (with correction for accidental coincidences) from $(31.0\pm1.7)$\% ($(28.0\pm1.5)$\% directly measured), to $(91.7\pm2.6)$\% ($(68.7\pm2.0)$\% directly measured) without central wavelength shift and up to $78$\% ($67$\% directly measured) with simultaneous $\pm1.1$~$\si{nm}$ spectral shift, which is far above the non-classical threshold of $50$\% \cite{kim2013conditions}. 

The presented scheme outperforms the standard filtering scheme due to higher transmission, limited by insertion loss only, and is expected to improve further \cite{he2020low,ranno2022integrated}. In addition, the development of low half-wave voltage EOPMs \cite{renaud2023sub} is expected to further improve the performance of the scheme. For now, the time lensing technique can be used for input bandwidths of at least tenths of GHz. Overcoming this challenge may require the development of more strongly dispersive media \cite{mazelanik2022optical} or a cascaded time lens system \cite{foster2009ultrafast}.

The measured non-classical visibility makes the presented technique promising for quantum communication and networking protocols such as entanglement swapping \cite{chou2007functional}, quantum teleportation \cite{bouwmeester1997experimental}. The presented solution offers efficient integration of spectrally mismatched quantum systems, making a step towards building hybrid quantum repeaters \cite{li2019experimental}, a vast hybrid quantum network \cite{Duan2001long} or realizing distributed quantum computing \cite{Main2015distributed} in a hybrid system. It opens new perspectives in photonic quantum computing by using novel photon sources as ancilla photons for heralded nonlinearities. One of the most promising examples is matching single photons produced via quantum dots to SPDC generated photon pairs, which currently can perform two-photon interference only with high-loss narrowband filtering \cite{huber2017interfacing,polyakov2011}. The time lensing technique presented in this work offers much higher power transmission than the filtering in that case, which directly converts to the success rate of the quantum communication protocols beyond quantum key distribution, distributed quantum computing and sensing.

This work was supported in part by the First TEAM programme of the Foundation for Polish Science (project no. POIR.04.04.00-00-5E00/18), co-financed by the European Union under the European Regional Development Fund and in parts by the National Science Centre (Poland) under the OPUS call within the Weave programme, project no. 2023/51/I/ST7/03068 and
under project no. 2022/45/N/ST2/04249, JS also acknowledges the support of START scholarship by the Foundation for Polish Science (FNP).

\nocite{kauffman1994time}
\nocite{drago2024hong}
\bibliographystyle{apsrev4-2.bst}
\bibliography{ref}

\appendix

\onecolumngrid
\newpage 
\setcounter{figure}{0}
\setcounter{equation}{0}

\begin{center}
    {\bfseries \Large Supplementary Material \\Quantum interference between spectral bandwidth mismatched photons}\\[0.7em]
    {\normalsize Jan Krzyżanowski,$^1$ Jerzy Szuniewicz,$^1$ Sanjay Kapoor,$^1$ Filip Sośnicki,$^{1,2}$ Michał Karpiński$^1$}\\ [0.5em]
    {\small $^1$Faculty of Physics, University of Warsaw, Pasteura 5, 02-093 Warszawa, Poland}\\
    {\small $^2$Integrated Quantum Optics, Institute for Photonic Quantum Systems (PhoQS), Paderborn University, Warburger Str. 100, D-33098, Paderborn, Germany}\\[0.5em]
\end{center}

\renewcommand{\theequation}{S.\arabic{equation}}
\renewcommand{\thefigure}{S\arabic{figure}}
\renewcommand{\bibnumfmt}[1]{[S#1]}
\renewcommand{\citenumfont}[1]{S#1}
\renewcommand{\thesection}{S\arabic{section}}

\title{Supplementary Material \\Quantum interference between spectral bandwidth mismatched photons}
\author{Jan Krzy\.zanowski}
\affiliation{Faculty of Physics, University of Warsaw, Pasteura 5, 02-093 Warszawa, Poland}
\author{Jerzy Szuniewicz}
\affiliation{Faculty of Physics, University of Warsaw, Pasteura 5, 02-093 Warszawa, Poland}
\author{Sanjay Kapoor}
\affiliation{Faculty of Physics, University of Warsaw, Pasteura 5, 02-093 Warszawa, Poland}
\author{Filip So\'snicki}
\affiliation{Faculty of Physics, University of Warsaw, Pasteura 5, 02-093 Warszawa, Poland}
\affiliation{Integrated Quantum Optics, Institute for Photonic Quantum Systems (PhoQS),Paderborn University, Warburger Str. 100, D-33098, Paderborn, Germany}
\author{Micha\l{} Karpi\'nski}
\affiliation{Faculty of Physics, University of Warsaw, Pasteura 5, 02-093 Warszawa, Poland}
\maketitle

\renewcommand{\theequation}{S.\arabic{equation}}
\renewcommand{\thefigure}{S\arabic{figure}}
\renewcommand{\bibnumfmt}[1]{[S#1]}
\renewcommand{\citenumfont}[1]{S#1}
\renewcommand{\thesection}{S\arabic{section}}

\section{Principle of operation of bandwidth converter}

\begin{figure}[!b]
\centering
\includegraphics[width=0.7\columnwidth]{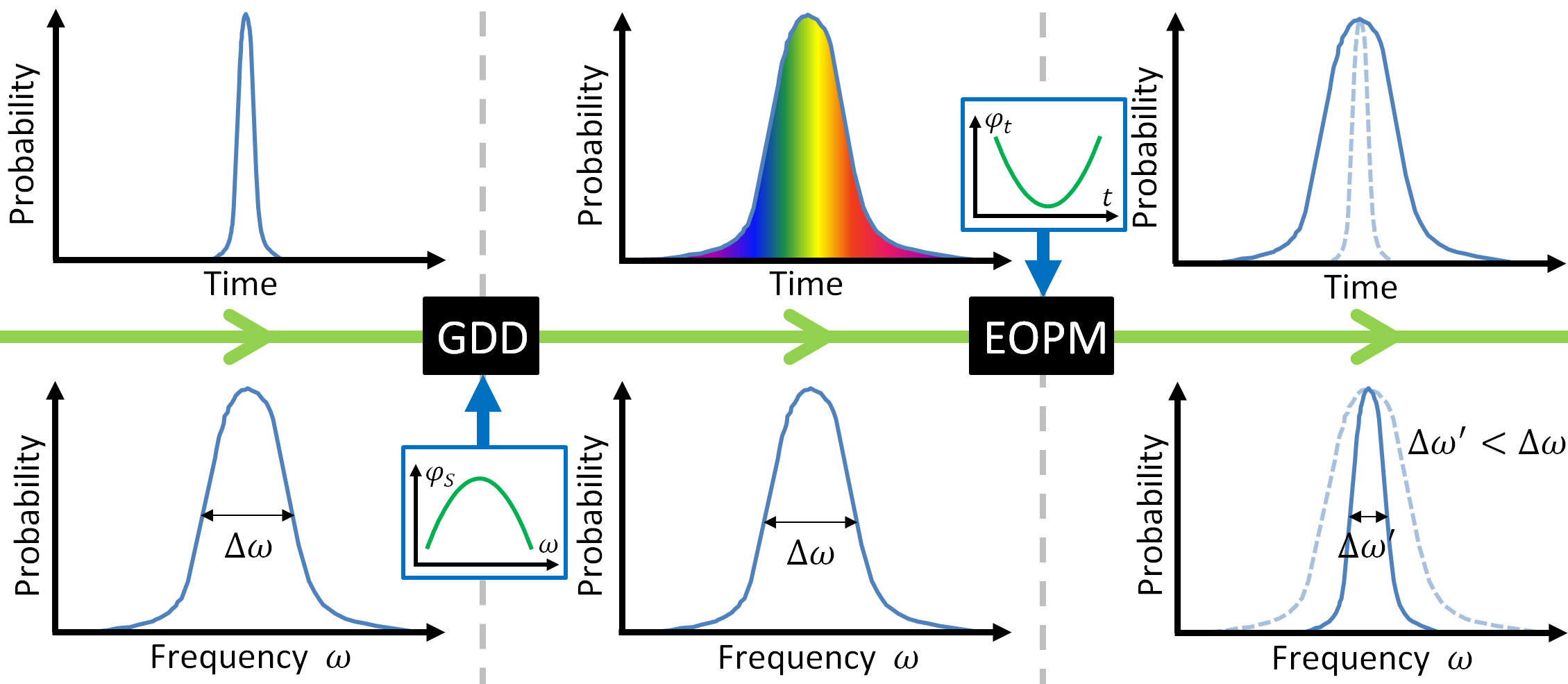}
\caption{The principle of operation of the spectral bandwidth converter. At the input the single-photon wavepacket (with approximately Gaussian envelope) is short in time and broad in spectrum (with optical frequency denoted by $\omega$ and spectral width denoted by $\Delta\omega$). The application of a quadratic phase in spectrum (denoted by $\phi_s$) using the dispersive medium (exhibiting group delay dispersion, GDD) broadens photon in time and chirps it. The chirp results in arrival of different frequency components (plotted with different colors within the stretched photon's temporal envelope) at different times. The photon is then subjected to quadratic temporal phase ($\phi_t$) modulation (in our experiment it is applied using an electro-optic phase modulator, EOPM). It shifts the spectral components such that they converge at the central frequency, resulting in spectral compression of the single-photon wavepacket (with spectral bandwidth denoted by $\Delta\omega'$).
\label{fig:TimeLens}}
\end{figure}

Here we discuss the principle of operation of the bandwidth converter. We show it schematically in Fig.~\ref{fig:TimeLens}. The spectral bandwidth compression requires manipulating the spectral and temporal phases of the light pulse. Since it features phase-only operations, in principle, it is intensity preserving. 

The first stage of the bandwidth converter is the application of a quadratic spectral phase to the  optical pulse, which chirps the pulse separating different spectral components linearly in time. The chirp also increases the temporal duration of the pulse. This part is experimentally realized using a dispersive medium with group delay dispersion (GDD), for example, a long single-mode fiber~[S1,~S2].

The second stage is the time lens, which applies a quadratic temporal phase. Since a linear temporal phase implies a Doppler shift, such a quadratic phase applies a Doppler shift that changes linearly in time. Modulating a previously spectrally chirped pulse in this manner allows to target pulse's different spectral components with different spectral shifts. If the spectral ($\Phi$) and temporal ($K_\text{eff}$) quadratic phase coefficients fulfill the time lens collimation condition~[S3,~S4]: 
\begin{equation}
    \Phi=\frac{1}{K_\text{eff}}, \label{eq:colimation}
\end{equation} one can obtain at the output an optical pulse with a different spectral bandwidth, which is the Fourier transform of the temporal profile and the temporal profile being the Fourier transform of the spectrum at the same time~[S3].

To experimentally apply the temporal phase, we used a traveling wave electro-optic phase modulator (EOPM), which is based on the Pockels effect. We used a single-tone RF signal to drive the modulator, obtaining a sinusoidal phase whose extrema can be approximated by a quadratic function~[S5]. To achieve the best compression factor, up to the aberrations introduced by the sinusoidal phase~[S4], the amplitude of the quadratic phase should satisfy the collimation condition, Eq.~\ref{eq:colimation}.

\section{Analytical model for ideal bandwidth converter}
In this section, we analytically derive the maximum Hong-Ou-Mandel (HOM) interference visibility attainable with an ideal bandwidth conversion setup operated at the time-lens collimation condition. Our approach to modeling the spectral bandwidth compression factor using Gaussian spectral wavepackets is based on the methodology outlined in the Supplementary Information of Ref.~[S1]. We consider two independent photons in distinct spatial modes labeled by $a$ and $b$.
\begin{align}
    \ket{\psi_a} &= \int \phi_a(\omega)\hat{a}^\dagger(\omega)\ket{\text{vac}},\\
    \ket{\psi_b} &= \int \phi_b(\omega)\hat{a}^\dagger(\omega)\ket{\text{vac}},
\end{align}
where $\hat{a}^\dagger(\omega)$ is the creation operator for a plane wave mode at frequency $\omega$. $\phi_{a(b)}$ are the Gaussian spectral amplitude functions defined as:
\begin{align}
    \phi_a(\omega) &= \frac{1}{(\pi \sigma_a^2)^{1/4}} \exp\left( - \frac{(\omega - \omega_0)^2}{2\sigma_a^2} \right) \label{eq:phi_a}, \\
    \phi_b(\omega) &= \frac{1}{(\pi \sigma_b^2)^{1/4}} \exp\left( - \frac{(\omega - \omega_0)^2}{2\sigma_b^2} \right) \label{eq:phi_b},
\end{align}
where $\sigma_a$ and $\sigma_b$ are the spectral bandwidth parameters. We assume a spectral bandwidth mismatch where $\sigma_a > \sigma_b$. To match the bandwidth of photon $a$ to photon $b$, we send the photon $a$ through a bandwidth conversion setup. This setup consists of a dispersive element with group delay dispersion (GDD) $\Phi$ followed by an ideal time lens (perfect quadratic temporal phase) with a chirp rate $K = 1/\Phi$.

The output spectral profile after bandwidth conversion can be determined using the time-to-frequency mapping proposed by Kauffman et al.~[S6].

First, we determine the temporal profile of photon $a$ via the Fourier transform of Eq. \eqref{eq:phi_a}. For a transform-limited Gaussian pulse, the temporal width $\sigma_{t,a}$ is inversely proportional to the spectral width $\sigma_a$. Specifically, the temporal envelope $E_a(t)$ is proportional to:
\begin{equation}
    E_a(t) \propto \exp\left( - \frac{t^2}{2 (1/\sigma_a)^2} \right) = \exp\left( - \frac{t^2 \sigma_a^2}{2} \right).
\end{equation}

The time-to-frequency mapping relates the output frequency axis $(\omega - \omega_0)$ to the time axis $t$ via the GDD para\-meter~$\Phi$:
\begin{equation}
    t = \Phi \times (\omega - \omega_0).
\end{equation}
Substituting this mapping into the temporal envelope gives the magnitude of the output spectral amplitude $\phi_a'(\omega)$:
\begin{equation}
    |\phi_a'(\omega)| \propto E_a(t)|_{t=\Phi(\omega-\omega_0)} \propto \exp\left( - \frac{\Phi^2 (\omega - \omega_0)^2 \sigma_a^2}{2} \right).
\end{equation}
We can define the new output bandwidth $\sigma_{a}'$ by inspecting the Gaussian exponent of the form $\exp(-(\omega-\omega_0)^2 / 2\sigma_{a}'^2)$:
\begin{equation}
\label{eq:newBandwidth}
    \frac{1}{2\sigma_{a}'^2} = \frac{\Phi^2 \sigma_a^2}{2} \implies \sigma_{a}' = \frac{1}{|\Phi| \sigma_a}.
\end{equation}
While the magnitude is transformed based on the input temporal shape, the time-lens setup imparts a residual quadratic spectral phase (chirp)~[S6]. The full complex expression for the transformed waveform is:
\begin{equation}
    \phi_a'(\omega) = \underbrace{\frac{1}{(\pi \sigma_{a}'^2)^{1/4}} \exp\left( - \frac{(\omega - \omega_0)^2}{2\sigma_{a}'^2} \right)}_{\text{magnitude}} \times \underbrace{\exp\left( i \frac{1}{2} \Phi (\omega - \omega_0)^2 \right)}_{\text{residual quadratic phase}}.
\end{equation}
We seek to compress the bandwidth of photon $a$ such that its output bandwidth matches photon $b$.
\begin{equation}
    \sigma_{a}' = \sigma_b.
\end{equation}
Using Eq.~(\ref{eq:newBandwidth}), we obtain:
\begin{equation}
    \frac{1}{|\Phi| \sigma_a} = \sigma_b.
\end{equation}
We can solve for the required GDD $\Phi$:
\begin{equation}
    \Phi = \frac{1}{\sigma_a \sigma_b}.
\end{equation}
This implies that the necessary dispersion is the inverse product of the input and target bandwidths.

\subsection{Coincidence probability calculations}
We now consider the interference of the transformed photon $a$ and the original photon $b$ at a $50:50$ beam splitter.

\begin{enumerate}
    \item \textbf{Mode $a$ (transformed):} With $\Phi = 1/(\sigma_a \sigma_b)$, the spectral width is equal to $\sigma_b$.
    \begin{equation}
        \phi_{a}'(\omega) = \frac{1}{(\pi \sigma_b^2)^{1/4}} \exp\left( - \frac{(\omega - \omega_0)^2}{2\sigma_b^2} \right) \exp\left( i \frac{(\omega - \omega_0)^2}{2\sigma_a \sigma_b} \right).
    \end{equation}
    \item \textbf{Mode $b$ (delayed):} Photon $b$ passes through a delay line introducing a time delay $\tau$.
    \begin{equation}
        \phi_{b, \tau}(\omega) = \phi_b(\omega) e^{-i\omega \tau} = \frac{1}{(\pi \sigma_b^2)^{1/4}} \exp\left( - \frac{(\omega - \omega_0)^2}{2\sigma_b^2} \right) e^{-i\omega \tau}.
    \end{equation}
\end{enumerate}

The coincidence probability $p(\tau)$ can be calculated using Eq. (17) from Drago et al.~[S7],
\begin{equation}
    p(\tau) = \frac{1}{2} - \frac{1}{2} \left| \int_{-\infty}^{\infty} d\omega \, \phi_{a}'(\omega) \, \phi_{b, \tau}^*(\omega) \right|^2.
\end{equation}

Substituting the specific functions and letting $\Omega = \omega - \omega_0$:
\begin{equation}
\label{eq:coincidenceInt}
    p(\tau) = \frac{1}{2} - \frac{1}{2} \frac{1}{\pi \sigma_b^2} \left| \int_{-\infty}^{\infty} d\Omega \, \exp\left( -\frac{\Omega^2}{\sigma_b^2} \right) \exp\left( i \frac{\Omega^2}{2\sigma_a \sigma_b} \right) \exp\left( i(\Omega + \omega_0)\tau \right) \right|^2.
\end{equation}

\subsection{Visibility bound set by the residual spectral phase}
To find the best possible minimum coincidence probability (maximum visibility), we set the time delay $\tau = 0$. The integral in Eq.~(\ref{eq:coincidenceInt}) becomes a Gaussian integral with a complex coefficient in the exponent.

Let the overlap integral be $J$:
\begin{equation}
    J = \frac{1}{\sqrt{\pi}\sigma_b} \int_{-\infty}^{\infty} d\Omega \, \exp\left[ -\Omega^2 \left( \frac{1}{\sigma_b^2} - i \frac{1}{2\sigma_a \sigma_b} \right) \right].
\end{equation}
Using the standard Gaussian integral identity $\int_{-\infty}^{\infty} e^{-Bx^2} dx = \sqrt{\frac{\pi}{B}}$, with $B = \frac{1}{\sigma_b^2} \left( 1 - i \frac{\sigma_b}{2\sigma_a} \right)$:
\begin{equation}
    J = \frac{1}{\sqrt{\pi}\sigma_b} \sqrt{\frac{\pi}{B}} = \frac{1}{\sigma_b \sqrt{B}} = \frac{1}{\sqrt{1 - i \frac{\sigma_b}{2\sigma_a}}}.
\end{equation}
The term required for the coincidence probability is the squared modulus $|J|^2$:
\begin{equation}
    |J|^2 = \frac{1}{\left| 1 - i \frac{\sigma_b}{2\sigma_a} \right|} = \frac{1}{\sqrt{1 + \left( \frac{\sigma_b}{2\sigma_a} \right)^2}}.
\end{equation}
Substituting $|J|^2$ back into the probability expression:
\begin{equation}
    p_{\text{min}} = \frac{1}{2} \left( 1 - \frac{1}{\sqrt{1 + \left( \frac{\sigma_b}{2\sigma_a} \right)^2}} \right).
\end{equation}
The maximum visibility of the interference can be calculated by the following expression~[S7]: 
\begin{equation}
{V_{\text{max}} = \frac{p_\text{max} - p_\text{min}}{p_\text{max}} = \frac{2F}{\sqrt{4F^2 + 1}},
\label{eq:vis}}
\end{equation}
where $F = \sigma_a/\sigma_b$ is the spectral bandwidth compression factor.

\begin{figure}[h]
\centering
\includegraphics[scale=0.75]{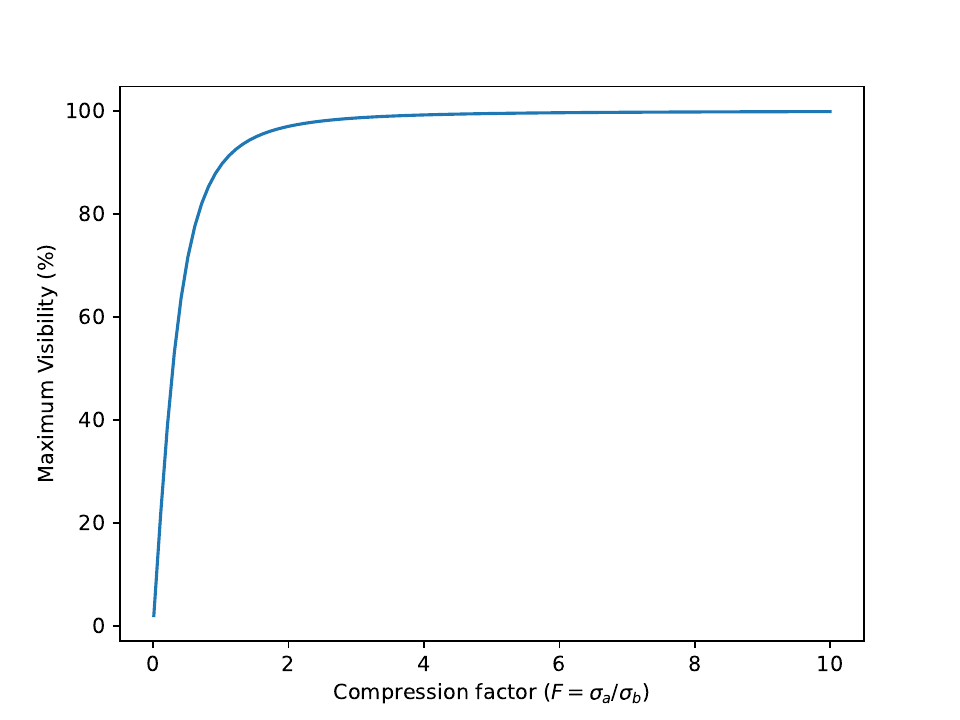}
\caption{Maximum visibility as a function of ratio of spectral bandwidth $\sigma_a$ of target photon to the bandwidth ($\sigma_b$) of input photon according to Eq.~(\ref{eq:vis}). As the compression factor increases, the maximum visibility approaches 100\%.}
\label{fig:S1}
\end{figure}

As shown in Fig.~\ref{fig:S1}, the visibility approaches 100\% as $F$ increases, because the residual phase variation over the target bandwidth becomes negligible. For instance, Eq.~\ref{eq:vis} gives $V_\text{max} = 97.01\%$ for a 2-fold compression, and $V_\text{max} = 99.87\%$ for a 10-fold compression. We emphasize that Eq.~\ref{eq:vis} is not an absolute upper bound on the achievable visibility, but rather the visibility obtained at the time-lens collimation condition $K=1/\Phi$, which is equivalent to collimating a diverging Gaussian beam by positioning a spatial lens at a distance equal to its focal length from the beam waist, which is valid in the regime where the focal length of the lens exceeds the Rayleigh range. In the temporal domain, this implies the length of the dispersion medium should be greater than the characteristic \emph{dispersion length}~[S3]. In the near-field, freeing $\Phi$ and $K$ as independent parameters allows the residual quadratic phase to be partially compensated, so the true optimum lies slightly above Eq.~\ref{eq:vis}. This is confirmed by the brute-force numerical optimization in Sec.~S3: for 6-fold compression Eq.~\ref{eq:vis} predicts $V=99.7\%$, whereas the optimization reaches $V=99.97\%$. Nonetheless, Eq.~\ref{eq:vis} provides an accurate closed-form estimate of the visibility achievable with an ideal time lens, capturing the essential physical trend that residual phase effects become negligible at higher compression factors.

\section{Numerical simulations of the sinusoidal time lens}

To model the experimental reality, we perform numerical simulations using a sinusoidal phase profile $\theta(t) = -A \cos(2\pi f_\text{m}t)$ provided by the electro-optic phase modulator. Here $f_\text{m}$ is the modulation frequency, and $A$ is the phase modulation amplitude. This profile is an approximation of the ideal parabolic phase profile, leading to higher-order temporal aberrations that degrade the HOM visibility~[S4]. We evaluate the bandwidth conversion performance across two distinct scenarios for a 6-fold spectral bandwidth compression of an input photon centered at $1547.5$~nm with a $1.78$~nm full-width at half-maximum (FWHM) bandwidth to a target of bandwidth $0.3$~nm. The results are summarized in Table~\ref{tab:sim_summary}.

\subsection{Numerical optimization for maximum visibility}
\begin{figure}[h]
    \centering
    \includegraphics[width=0.9\linewidth]{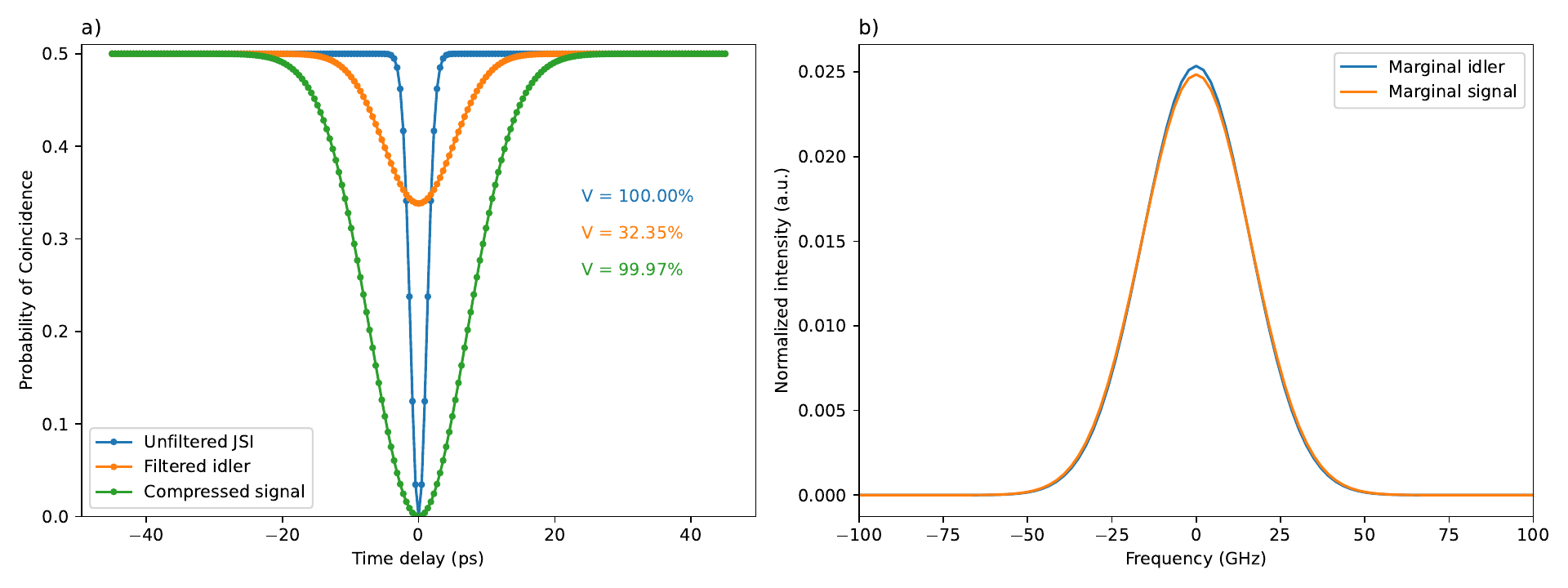} 
    \caption{Spectral bandwidth compression with numerically optimized parameters. (a) Simulated Hong-Ou-Mandel (HOM) interference for the initial biphoton state (blue), the state after filtering the idler (orange), and the state after signal compression using optimized sinusoidal time lens parameters (green). Brute-force optimization of the modulation frequency, amplitude, and GDD allows for the retrieval of near-perfect indistinguishability ($V = 99.97\%$). (b) Marginal spectral distributions of the filtered idler and the compressed signal. Under these optimal parameters, the compressed signal closely matches the Gaussian profile of the target idler.}
    \label{fig:S2}
\end{figure}

This simulation serves to establish the highest possible HOM visibility achievable using a sinusoidal time-lens-based bandwidth converter, by performing a brute-force optimization of all system parameters: GDD ($\Phi$), modulation frequency ($f_m$), and modulation amplitude ($A$). We found the optimal set of parameters to be $f_m = 4.1\text{ GHz}$, $A = 57.11\pi$ rad, and $\Phi = 8.23\text{ ps}^2$, to match input bandwidth of 1.78 nm to 0.3 nm. This configuration results in a maximum theoretical visibility of $V = 99.97\%$ as shown in Fig.~\ref{fig:S2} (a). The corresponding marginal spectral distributions of the photons are shown in Fig.~\ref{fig:S2} (b). The converted (signal) photon is nearly perfectly matched with the target spectral profile of the idler photon.

\subsection{Simulation of experimental configuration}
Finally, we simulate the HOM interference with specific parameters used in the experimental setup described in the main text while incorporating uncertainties, to validate the measured visibility. The simulation does not include the optical losses or accidental coincidences. It therefore predicts the intrinsic visibility set by the spectral-mode overlap of the bandwidth converted and reference photon, which is the quantity that should be compared with the accidental-corrected experimental values.

\begin{figure}[h]
    \centering
    \includegraphics[width=0.95\linewidth]{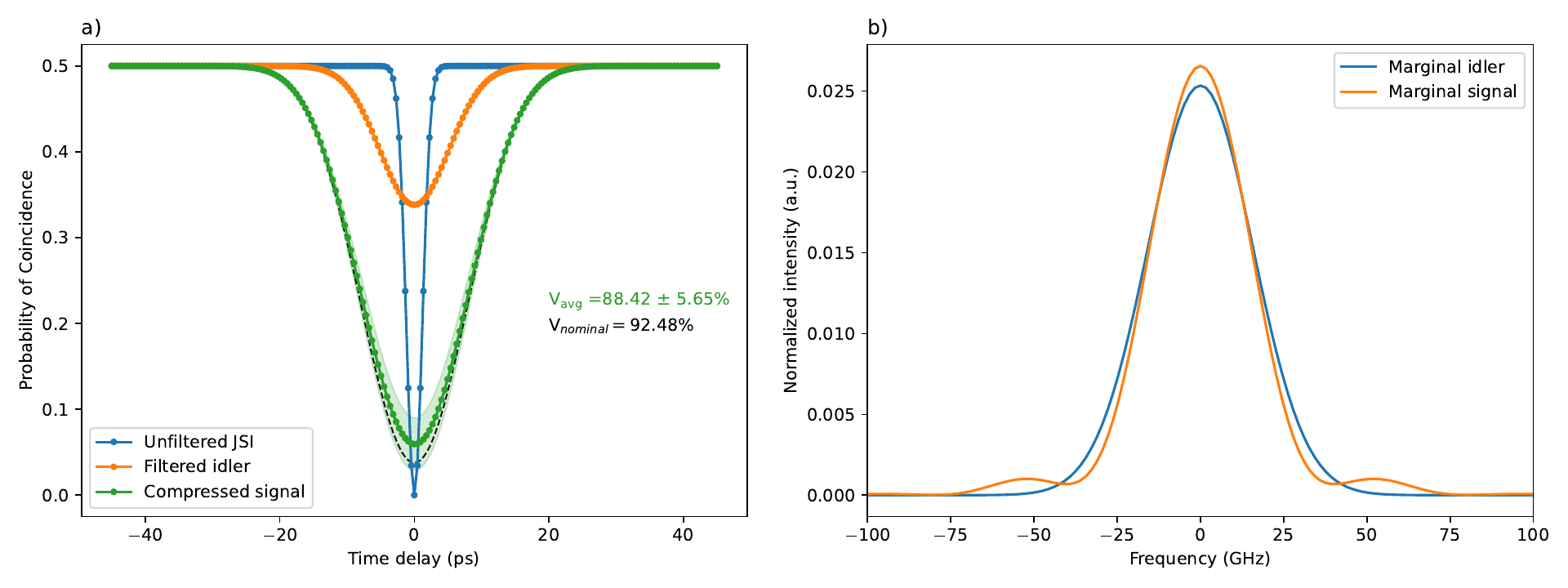}
    \caption{Simulation of the laboratory experimental configuration. (a) HOM interference simulated using the parameters employed in the experiment while accounting for uncertainty ($\Phi = 10.7\pm0.1$ ps$^2$, $f_\text{m} = 16$~GHz, $A = 11.31\pm0.31$ rad). The simulation also incorporates a $\pm1$~ps uncertainty in the temporal alignment of the signal photon being compressed and the time lens. The resulting average visibility of $(88.4\pm5.6)\%$ which is in agreement with the experimentally measured value of $(91.2\pm1.4)$\%, after correcting for accidentals. (b) Marginal spectral distributions showing prominent side-lobes in the signal photon for the best-case scenario, where the visibility reaches 92.5\%.}
    \label{fig:S4}
\end{figure}

We use the GDD of the 500-m-long SMF-28 fiber spool, $\Phi = 10.7\pm0.1\text{ ps}^2$, the modulation frequency at $f_m = 16\text{ GHz}$, and the modulation amplitude of $A = (3.6\pm0.1)\pi$ rad measured using electro-optic side-band analysis. The simulations also include a $\pm 1$~ps temporal fluctuation between the time lens and the photons. At the nominal parameters the visibility is limited to 92.5\% by the sinusoidal aberrations combined with the non-optimal dispersion (Fig.S4(b)). Averaging over the parameter fluctuations yields $V = (88.4\pm5.6)\%$, see Fig.~\ref{fig:S4} (a). The marginal distribution for the nominal parameter values (Fig.~\ref{fig:S4} (b) shows the more pronounced spectral side-lobes compared to Fig.~\ref{fig:S2} (b), limiting the maximum visibility to 92.5\%. The intrinsic simulated visibility agrees with the accidental-corrected experimental value of $(91.7\pm2.6\%$). The lower measured raw visibility of $(68.7 \pm 2.0)$ is accounted for by accidental coincidences, which arise from the imbalanced photon number at the two beam-splitter inputs and the asymmetric losses between the interferometer arms. Accidental coincidences are thus the dominant limitation of the raw measured visibility, whereas the intrinsic performance of the converter is primarily limited only by the time-lens aberrations.

\begin{table}[h]
\centering
\caption{Comparison of simulated maximum visibility cases for sinusoidal time lens}
\label{tab:sim_summary}
\begin{tabular}{@{}llccccl@{}}
\toprule
Case & Description & $f_\text{m}$ (GHz) & $A$ ($\pi$ rad) & GDD ($\text{ps}^2$) & Visibility \\ 

1 & Numerical optimization & 4.1 & 57.11 & 8.23 & 99.97\% \\

2 & Experimental configuration & 16.0 & $3.6\pm0.1$ & $10.7\pm0.1$ & $(88.4\pm5.6)$\% \\ 

\end{tabular}
\end{table}

\section{Simultaneous spectral shift and bandwidth manipulation}
\begin{figure}[h]
    \centering
    \includegraphics[width=0.95\linewidth]{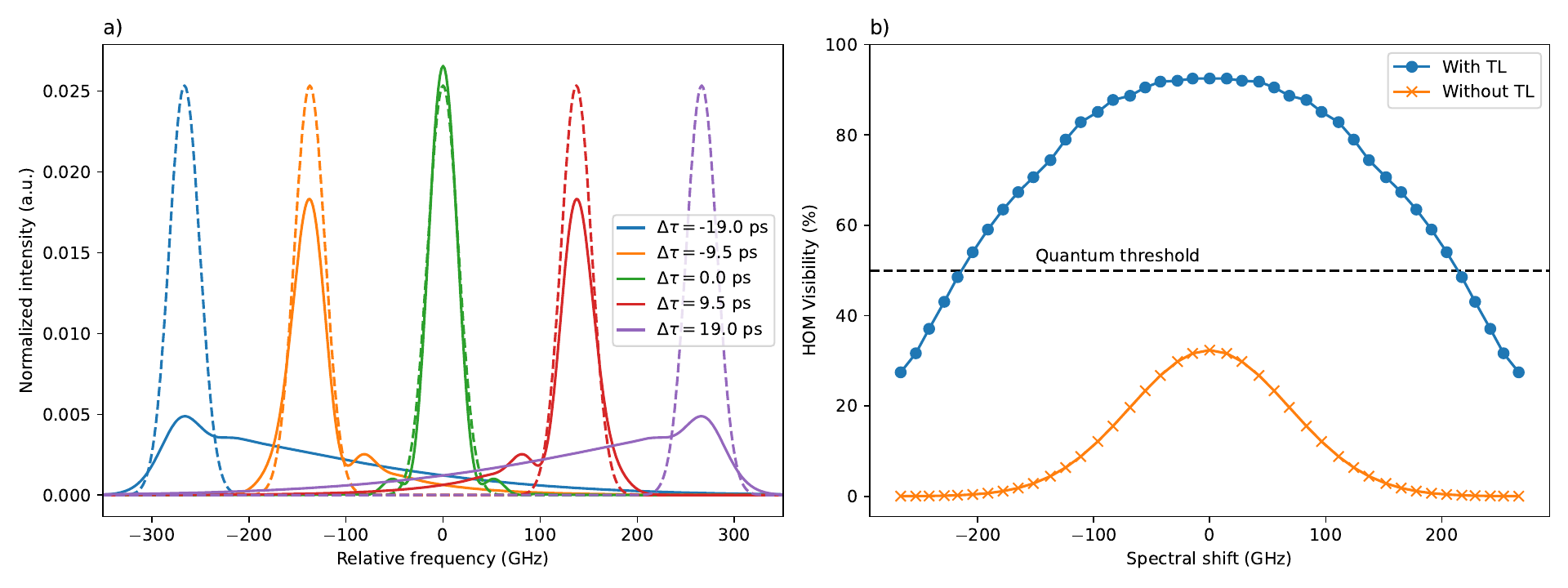}
    \caption{Simultaneous spectral shift and bandwidth compression. (a) Simulated marginal distributions of the compressed signal photon for various temporal delays $\Delta \tau$ relative to the time lens. Solid lines represent the compressed signal with various magnitudes of spectral shift, while dashed lines indicate the corresponding ideal Gaussian target spectral profiles. (b) The visibility is highest when the shift is zero and decreases as the signal photon experiences higher-order aberrations from the asymmetric sinusoidal phase profile at larger delays.}
    \label{fig:S5}
\end{figure}

By introducing a precise temporal delay $\Delta \tau$ between the signal photon and the peak of the sinusoidal phase modulation, we can achieve spectral shifting and bandwidth compression simultaneously, a process analogous to how a spatial lens shifts an off-axis beam passing through it~[S8]. In a time-lens setup, the spectral shift is determined by the temporal gradient of the phase imparted by the electro-optic modulator. As shown in Fig.~\ref{fig:S5}(a), varying $\Delta \tau$ allows for a tunable shift of the signal photon's central frequency. However, moving away from the extremum of the sinusoidal drive exposes the photon to non-parabolic regions of the phase profile, introducing temporal aberrations, which degrade the spectral overlap with the target idler photon spectral profile. Consequently, as quantified in Fig.~\ref{fig:S5}(b), the maximum achievable Hong-Ou-Mandel visibility is highest at zero shift and gradually decreases as the magnitude of the spectral shift increases.

\bigskip
\bigskip

\section{Experimental details and data}
\subsection{Joint spectral intensities of the modified photons}
\begin{figure}[h]
    \centering
    \includegraphics[width=1.0\linewidth]{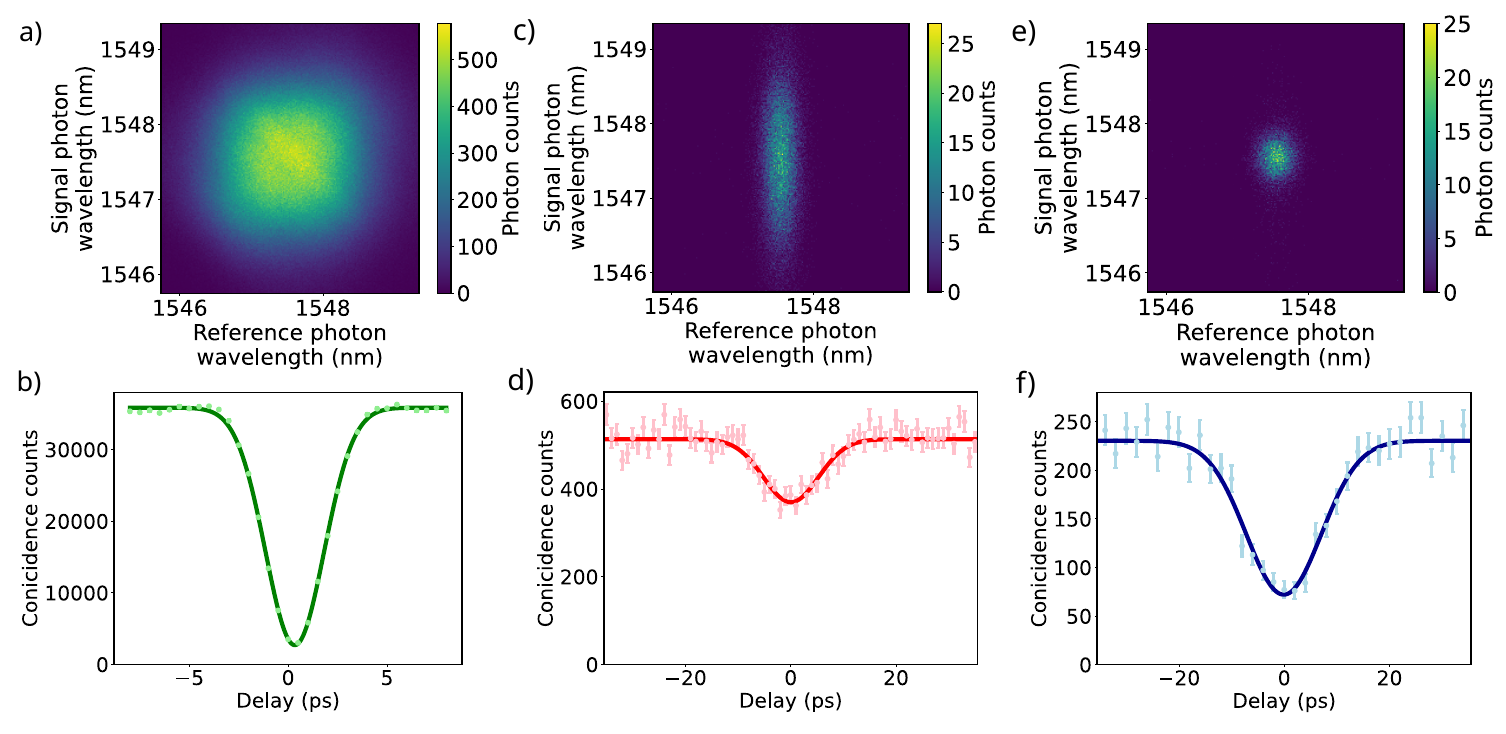}
    \caption{Joint spectral intensities of signal and reference photons: (a)~without any modification, (c)~with reference photon filtered and signal photon unmodified, (e) signal photon compressed together with filtered reference photon. The JSIs are measured with DFT technique. The reference photon is filtered with the spectral filter centered at the input spectrum. (b),~(d),~(f): the HOM dips corresponding to the JSIs in the upper panels, data points: measured experimental data, solid lines: Gaussian fit. Error bars assume Poisson photon number statistics.\label{fig:S7}}
\end{figure}

We show the measured joint spectral intensities (JSI) of the photons without any modification (Fig.~\ref{fig:S7}(a)), with the filter applied to the reference photon (Fig.~S6(c)) and after application of the bandwidth compression to the signal photon (Fig.~S6(e)). All JSIs are measured using the DFT technique with $40.9$~$\si{km}$ and $41.8$~$\si{km}$ of SMF-28 fiber as a dispersive medium for each photon. The reference photon is filtered with the spectral filter centered at the input spectrum. The shape of the JSIs shows that there is no spectral correlation in the JSI before introducing the mismatch (Fig.~\ref{fig:S7}~(a)). In (Fig.~\ref{fig:S7}~(b))  we show the HOM dip for the unmodified photons; the interference visibility is $(92.5\pm0.1)$\%. The narrowband spectral filter added in Fig.~\ref{fig:S7}~(c) and~(d) introduces significant distinguishability between signal and reference photons, which is confirmed by the asymmetric JSI shape and the directly measured HOM visibility of $(28.0\pm1.5)$\%. The bandwidth converter added in Fig.~\ref{fig:S7}~(e) and~(f) compresses the signal photon restoring the indistinguishability. It results in symmetric JSI again, along with directly measured HOM visibility of $(68.7\pm2.0)$\%. The same data as in Fig.~\ref{fig:S7}~(d) and~(f) we show in Fig.~3(a) in the main text.

\subsection{Hong-Ou-Mandel measurements}
\subsubsection*{Raw data}
\begin{figure}[h]
    \centering
    \includegraphics[width=1.0\linewidth]{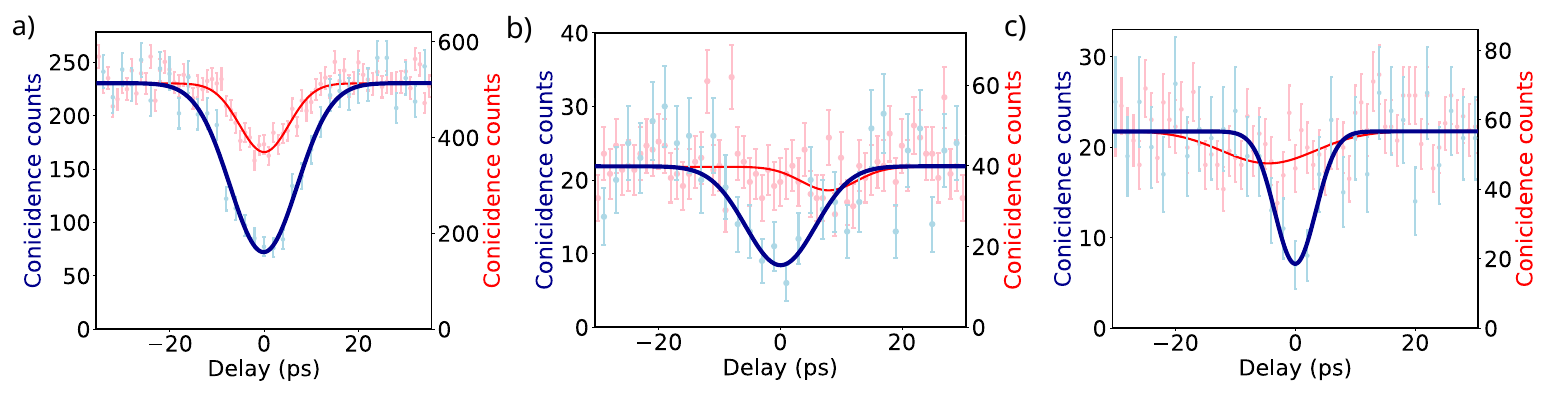}
    \caption{Hong-Ou-Mandel dips without normalization. In red we plot the dips measured without spectral compression and in blue with the converter matching the photons' spectra. The data points represent raw experimental coincidence counts. Error bars assume Poisson photon count statistics. With solid lines we plot the Gaussian fits to the corresponding data with the same parameters as in Fig.~3 in the main text. In (a) we show the dips with the filter, applied to the reference photon, centered at its spectrum, whereas in~(b) and~(c) the filter is blue and red shifted, respectively, by $1.1$~$\si{nm}$.
    \label{fig:S8}}
\end{figure}

In Fig.~\ref{fig:S8}, we provide plots of directly measured coincidence counts for HOM interference without any normalization or correction. These are the raw experimental data used to produce the  Fig.~3~(a,c,e) in the main text but without normalization. In Fig.~\ref{fig:S8}~(a) we show the HOM interference data for the reference photon filtered with the filter centered on its spectrum. In Fig.~\ref{fig:S8}~(b,c) we show the HOM interference results with the filter applied to the reference photon shifted by $1.1$~$\si{nm}$ from its initial position. For the error bars we assume Poisson photon count statistics.

\subsubsection*{Accidental coincidences correction}

The HOM interference is observed in the number of coincidence events between the two outputs of the 50:50 beam splitter (BS) measured as a funcion of delay $\tau$. In principle there are two possible ways of recording a coincidence event at BS: one photon comes from the signal path, one comes from the reference path ($C(\tau)$), interfering at the BS, or a pair of photons coming from either signal or reference path ($C_{\mathrm{acc}}$) is split at the BS into two output ports. Therefore, the measured number of coincidence events in HOM experiment is equal to:

\begin{equation}
    C_{\mathrm{meas}}(\tau) = C(\tau) + C_{\mathrm{acc}}.
\end{equation}

Whereas the coincidence events of the first type, $C(\tau)$, correspond exactly to the results of two-photon interference, the second type, ($C_{\mathrm{acc}}$), called accidental coincidences, are experimental imperfections. A pair of photons in one path appears in our experiment due to the finite extinction ratio of the FPBS used to split the photon pairs generated in type-II SPDC process. The photons after splitting traverse a number of optical elements introducing different losses, making the contribution of accidental coincidences non-negligible.

Using the conventional definition of HOMI visbility $V=(C_{\mathrm{max}}-C_{\mathrm{min}})/C_{\mathrm{max}}$ the accidental terms cancel out in the numerator but not in the denominator, yielding
\begin{equation}
    V_{\mathrm{meas}} = \frac{V_0}{1+C_{\mathrm{acc}}/C(\tau\!\to\!\infty)}
    = V_0\frac{1}{1+1/\mathrm{CAR}},
    \label{eq:vis_acc}
\end{equation}
where $V_0$ is the HOMI visibility without accidental coincidences and $\mathrm{CAR}=C_0/C_{\mathrm{acc}}$ is the coincidence-to-accidental ratio. If $\mathrm{CAR}$ is comparable to $1$ (as can be the case in our experiments) the correction to the HOMI visibility becomes non-negligible.

To mitigate that problem we independently measured the rates of accidental coincidences. We disconnected the reference photon from the Hong-Ou-Mandel interference beam splitter and measured the rate of coincidences arising from the signal photon's path only. We repeated the procedure for the reference photon for the measurement without the bandwidth converter. We experimentally verified that there was no significant difference in the number of accidental coincidences for different delays. We summarize the measured accidental coincidence rates in Table~\ref{tab:Accidentals}. To correct the measured HOM dip coincidence counts we subtracted the number of accidental coincidence derived from the rates shown in Table~\ref{tab:Accidentals} from the raw HOMI data shown in Fig.~\ref{fig:S8} obtaining the corrected data shown in Fig.~3(a,c,e). After the subtraction, we obtain
\begin{equation}
    C_{\mathrm{correceted}}(\tau) = C_{\mathrm{meas}}(\tau) - C_{\mathrm{acc}} = C(\tau),
\end{equation}
which contains only the result of HOMI, without the contribution of accidental coincidences.

\begin{table}[h]
\centering
\caption{Averaged accidental coincidence counts, $30$~$\si{s}$ acquisition time.}
\label{tab:Accidentals}

\begin{tabular}{@{}l|cc|c|cc|c|cc|c@{}}
\toprule

Filter shift & No & No & No & Blue shift & Blue shift & Blue shift & Red shift & Red shift & Red shift \\ 

Photon's path & Signal & Reference & Signal & Signal & Reference & Signal & Signal & Reference & Signal \\ 

Converter in the setup & No & No & Yes & No & No & Yes & No & No & Yes \\ 

Accidental counts/$30$~$\si{s}$ & $25.6\pm0.8$ & $24.8\pm0.8$ & $57.6\pm0.8$ & $2.6\pm0.4$ & $3.2\pm0.4$ & $3.3\pm0.3$ & $3.9\pm0.3$ & $1.6\pm0.2$ & $3.0\pm1.7$ \\

\end{tabular}

\end{table}

\subsection{Confirmation of zero delay position for low visibility dips}

\begin{figure}[h]
    \centering
    \includegraphics[width=0.7\linewidth]{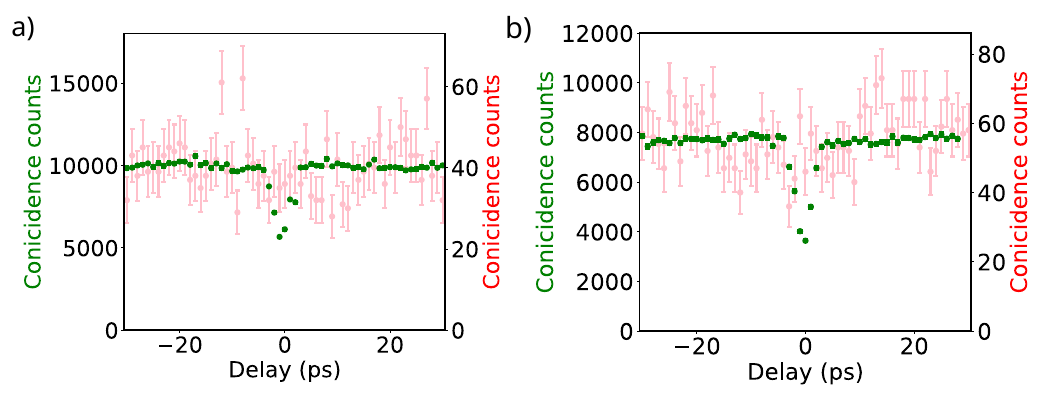}
    \caption{Hong-Ou-Mandel dips used to determine the zero delay for the measurements without spectral modification and shifted filter. We show in red the same dips without spectral modification as in Fig.~3(c,e) in red, for the blue shifted filter~(a) and red shifted photon~(b), along with the dips without any spectral modification (neither filtering or compression) applied to any of the photons. The dips were acquired in the same experimental configuration, unlike the dips shown in Fig.~3. This allowed us to confirm the correctness of the delay used to acquire the red data, set the zero point at the delay axis for it, and plot the data in Fig.~\ref{fig:S8} and Fig.~3 in the main text using a common delay axis. Error bars assume Poisson photon count statistics, coincidences were acquired for $30$~$\si{s}$ at each point}
    
    \label{fig:S9}
\end{figure}

The red and blue data in Fig.~3(a,c,e) and Fig.~\ref{fig:S8} were acquired in different experimental configurations. For measuring the dip without spectral compression, we removed all the components required for the demonstration of spectral bandwidth modification (marked with asterisks in Fig. 2(a)), to avoid introducing additional GDD and, therefore, distinguishability between the photons. This different experimental configuration required finding the temporal location of the dip independently. While in Fig.~3(a) the location of the dip is clearly visible, the location of the HOM dips for the mismatched photons with shifted central wavelength from Fig.~3(c,e) is not obvious. Since the direct determination of the zero delay based on the minimum of the dip itself was unreliable, we used an alternative method to find it.

We verified the position of the dip for the mismatched photons by performing another HOM measurement without spectrally  filtering the reference photon. We opened the slit in the $4f$ filter to disable the filtering without modifying the delay. We present the results of that measurement in Fig.~\ref{fig:S9} in green. For the green dip, none of the photons has undergone bandwidth conversion or spectral filtering. We set the minimum of that dip as zero delay reference for the horizontal axis of Fig.~\ref{fig:S9}. The accidental coincidence rate for the green dips was significant, reducing the visibility of the interference (which we do not correct). This arises from the attenuation in the signal photon path, which needs to be in place to keep the path length the same as for the target measurement.

Subsequently, we applied the shifted narrow spectral filter by closing the slit in the $4f$ filter. 
No other modifications were made to the optical system. We then acquired the HOM interference data between the bandwidth mismatched photons, shown in red in Fig.~3(c,e) and~\ref{fig:S8}(b,c).  The clearly visible dip in green shows that the red data was taken at the correct delay. The measurements from Fig.~\ref{fig:S9} together with the clearly visible dip with spectral modification (blue data in Fig.~\ref{fig:S8}(b,c) and in Fig.~3(c,e)), allowed us to plot the red and blue data in Fig.~\ref{fig:S8}(b,c) and in Fig.~3(c,e) using a common delay axis.

\bigskip

[S1] M. Karpński, M. Jachura, L. J. Wright, and B. J. Smith, Nature Photon. 11, 53 (2017).

[S2] F. Sośnicki, M. Mikolajczyk, A. Golestani, and M. Karpiński, Appl. Phys. Lett. 116, 234003 (2020).

[S3] V. Torres-Company, J. Lancis, and P. Andres, Prog. Opt. 56, 1 (2011).

[S4] S. Kapoor, F. Sośnicki, and M. Karpi´nski, APL Photonics 10, 096111 (2025).

[S5] B. H. Kolner, Appl. Phys. Lett. 52, 1122 (1988).

[S6] M. Kauffman, W. Banyai, A. Godil, and D. Bloom, Appl. Phys. Lett 64, 270 (1994).

[S7] C. Drago and A. M. Brańczyk, Can. J. Phys. 102, 411 (2024).

[S8] L. J. Wright, M. Karpiński, C. S¨oller, and B. J. Smith, Phys. Rev. Lett. 118, 023601 (2017).

\end{document}